# Cavity-enhanced zero-phonon emission from an ensemble of G centers in a silicon-on-insulator microring

B. Lefaucher[1], J.-B. Jager[1], V. Calvo[1], A. Durand[2], Y. Baron[2], F. Cache[2], V. Jacques[2], I. Robert-Philip[2], G. Cassabois[2], T. Herzig[3], J. Meijer[3], S. Pezzagna[3], M. Khoury[4], M. Abbarchi[4], A. Dréau[2], J.-M. Gérard[1,a].

[1] *Université Grenoble Alpes, CEA, Grenoble INP, IRIG, PHELIQS, Grenoble, France.*

[2] *Laboratoire Charles Coulomb, Université de Montpellier and CNRS, Montpellier, France.*

[3] *Division of Applied Quantum Systems, Felix-Bloch Institute for Solid-State Physics, University Leipzig, Leipzig, Germany.*

[4] *CNRS, Aix-Marseille Université, Centrale Marseille, IM2NP, UMR 7334, Marseille, France.*

a) Author to whom correspondence should be addressed: jean-michel.gerard@cea.fr

ABSTRACT

We report successful incorporation of an ensemble of G centers in silicon-on-insulator (SOI) microrings using ion implantation and conventional nanofabrication. The coupling between the emitters and the resonant modes of the microrings is studied using continuous-wave and time-resolved microphotoluminescence (PL) experiments. We observe the resonant modes of the microrings on PL spectra, on the wide spectral range that is covered by G centers emission. By finely tuning the size of the microrings, we match their zero-phonon line at 1278 nm with a resonant mode of quality factor around 3000 and volume 7.2 $(\lambda/n)^3$. The zero-phonon line intensity is enhanced by a factor of 5, both in continuous-wave and time-resolved measurements. This is attributed to the Purcell enhancement of zero-phonon spontaneous emission into the resonant mode and quantitatively understood considering the distribution of the G centers dipoles. Despite the enhancement of the zero-phonon emission, we do not observe any sizeable decrease of the average lifetime of the G centers, which points at a low radiative yield (<10%). We reveal the detrimental impact of parasitic defects in heavily implanted silicon, and discuss the perspectives for quantum electrodynamics experiments with individual color centers in lightly implanted SOI rings. Our results provide key information for the development of deterministic single photon sources for integrated quantum photonics.

Silicon-on-insulator (SOI) technology is a promising platform for the development of large-scale integrated quantum photonics (IQP) due to its compatibility with mature microelectronics[1,2]. The integration of four-wave mixing photon-pair sources in silicon photonic circuits has already enabled key results in IQP, such as quantum interference and multi-dimensional entanglement[3]. However, no deterministic single-photon sources, that emit on-demand exactly one photon in response to a trigger signal, have been demonstrated in silicon. Such emitters are necessary for applications requiring on-demand generation of single-photon states with high efficiency, such as quantum computing and quantum repeaters[4]. In that regard, silicon lags far behind other platforms such as III-V semiconductors in which nearly ideal single-photon emission has been achieved.

The recent isolation of artificial atoms in silicon emitting at telecom wavelengths, such as the G center[5], the T center[6] and other carbon-related defects[7–9], and the W center[10] provides a route toward on-chip deterministic single-photon emission. Ensembles of G centers in bulk silicon and planar SOI have already been extensively studied[11–14], which was motivated by their sharp Zero-Phonon Line (ZPL) at 1278 nm matching the telecom O-band. The introduction of ensembles of G centers in nanopatterned SOI has been explored in the quest for lasing[15] as well as for enhancing the extraction of their emission out of the Si matrix[16]. In this letter, we report successful incorporation of an ensemble of G centers into SOI microrings using ion implantation and standard nanofabrication, and demonstrate cavity-enhanced zero-phonon spontaneous emission (SE) into a resonant mode.

The sample was fabricated from a commercial SOI wafer from SOITEC with 220 nm silicon on 2.5 μm buried oxide. A 1 cm² square piece from the wafer was implanted with carbon atoms with a dose of $5\times10^{14}$ at/cm² at 30 keV and 0° angle. The sample was then flash-annealed at 1000°C for 20s in $N_2$ atmosphere and irradiated with 95 keV protons with a fluence of $5\times10^{13}$ at/cm². This procedure leads to the creation of dense ensembles of G centers. The microrings were fabricated using a standard combination of electron-beam lithography and reactive ion etching, with a width of 400 nm and an average diameter ranging from 2.4 to 2.6 µm in 5 nm steps. In addition, each microring features an external ring of 200 nm width with a 500 nm gap. The external ring aims at scattering photons escaping from the equatorial plane of the microring towards the objective to improve the collection efficiency at normal incidence. A scanning electron microscope image of a typical microring is shown in Fig. 1(a). The same set of microrings was fabricated in a virgin SOI layer to verify that the nanofabrication process is not responsible for the creation of G centers. This is an important point as a carbonaceous gas ($CH_2F_2$) was used in the reactive ion etching step. For example, the exposure to carbon-tetrafluoride ($CF_4$) has been reported to introduce G centers in silicon thin films[17]. Nevertheless, no photoluminescence (PL) was detected from this reference set of microrings.

The sample was held at 30 K on the cold finger of a closed-cycle cryostat. Optical excitation was provided by a continuous-wave (cw) laser diode or a pulsed laser diode, both emitting at 532 nm. The laser beam was focused onto the sample at normal incidence using a microscope objective of numerical aperture NA=0.85 installed inside the cryostat. The PL of the sample was collected using the same objective and analyzed using a grating monochromator and an InGaAs linear photodiode array camera. The spectral resolution is around 0.2 nm. For time-resolved experiments, the PL signal was measured by a fiber-coupled superconducting nanowire single photon detector. The PL maps were obtained using a steering mirror to move the excitation spot across the sample. The spatial resolution of the maps is approximately 1 µm, corresponding to the diffraction limit of the objective in the near-infrared. The PL signal within the 1250 nm to 1450 nm spectral range was selected using a band-pass filter and measured in a continuous way during the scanning.

We will focus our attention on a set of five microrings labelled R1 to R5, with respective diameters ranging from 2.415 to 2.435 µm. Fig. 1(b) shows a PL intensity map under cw excitation at 10 µW (~1 kW/cm² incident power on the sample), in the linear power regime of the emitters. The observation of bright PL rings show that optically active defects are incorporated in the cavities. A high-resolution PL map of R3 is shown in Fig. 1(c). The bright internal ring shows that the concentration of emitters is rather uniform. The purple halo with lower PL intensity is attributed to emission from the external ring.

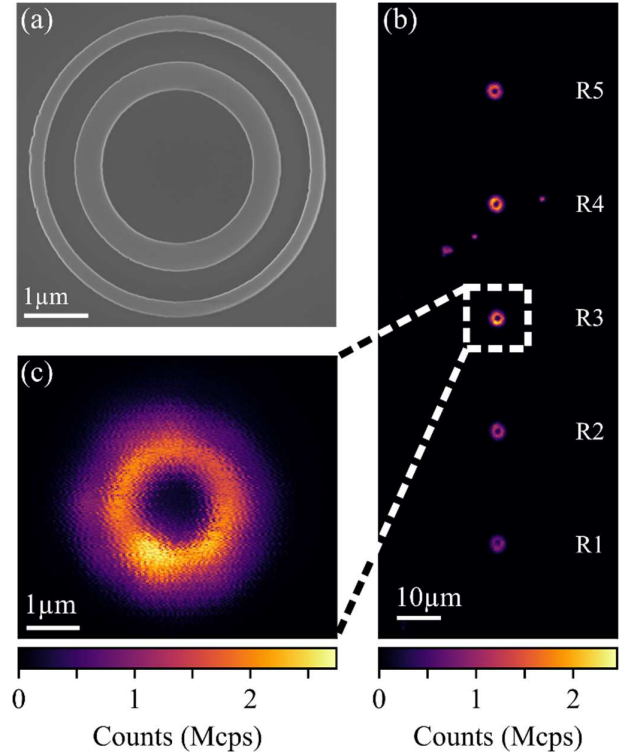

FIG. 1. SOI microrings containing G centers. (a) Scanning electron microscope image of a microring of diameter 2.8 µm and width 400 nm. The external ring is 500 nm apart from the microring and has a width of 200 nm. (b) PL map over five microrings with average diameters ranging from 2.415µm (R1) to 2.435 µm (R5) with 5 nm steps. (c) High-resolution PL map of R3. The PL maps were recorded at 30 K.

A PL spectrum of a 2 µm thick waveguide fabricated in the same sample and serving as a reference is shown in Fig. 2(a). It is identical to prior results for ensembles of G centers at cryogenic temperature, with a ZPL at 1278 nm, a broad Phonon Side Band (PSB) and the 'E line' at 1380 nm, with a Debye-Waller factor $\xi \approx 15\%$[11,13,18]. The ZPL fits to a Gaussian profile with a full width at half maximum (FWHM) of 1.1 ± 0.1 nm, which is attributed to inhomogeneous broadening.

As shown in Fig. 2(b), the spectra display the resonant modes of the microrings on the PSB. The free spectral range (FSR) between the modes is equal to 35.6 ± 0.1 meV ($\approx$ 52.3 nm ± 0.2 nm at 1350 nm) for R3. From finite-element simulations using the FEMSIM software from SYNOPSYS, we calculated the intensity profile and effective index of the $TE_{00}$ mode of a curved waveguide, whose cross-section and radius of curvature are identical to the ones of R3, for wavelengths between 1.2 and 1.5 µm (see the supplementary information). A group index of 4.53 was calculated at 1.35 µm. This leads to an estimate of the FSR around 52.8 nm, in close agreement with the experimental value. We conclude that all resonances on the spectra correspond to a single family of $TE_{00,m}$ resonant modes engendered by the $TE_{00}$ guided mode of the waveguide. The simulations permit identifying without ambiguity the azimuthal quantum number m of each resonant mode, used as label in Fig. 2(b).

Quality factors (Q) of typically 3100 ± 200 were measured for all the modes on high-resolution spectra, as shown in Fig. 2(c). For such small diameter rings, one expects Q to be either limited by intrinsic radiation losses or by scattering by sidewall roughness, depending on the mode wavelength[19]. For ring R3, 3D FDTD simulations show that bending losses correspond to Q around 7500 at 1.45 µm and 76000 at 1.28 µm. Additionally, propagation losses in dry-etched SOI waveguides typically range between 7 and 3 dB/cm, which corresponds to scattering-limited Qs in the [70000-120000] range[20,21]. We conclude that neither intrinsic bend losses nor scattering by side-wall roughness can account for the low Q of the modes of our rings. Furthermore, the G centers PSB is not expected to induce anti-Stokes absorption for the resonant modes at low temperature due to the small phonon population. For these reasons, we conclude that Q is limited by absorption losses due to parasitic defects (such as vacancies[22]) that have been formed besides G centers during fabrication. Besides, the presence of parasitic emitters – of unknown nature at this stage – is revealed at higher excitation power by the observation of the $TE_{00,16}$ mode (~1233 nm) that cannot be excited by the G centers (see the supplementary Fig. S2).

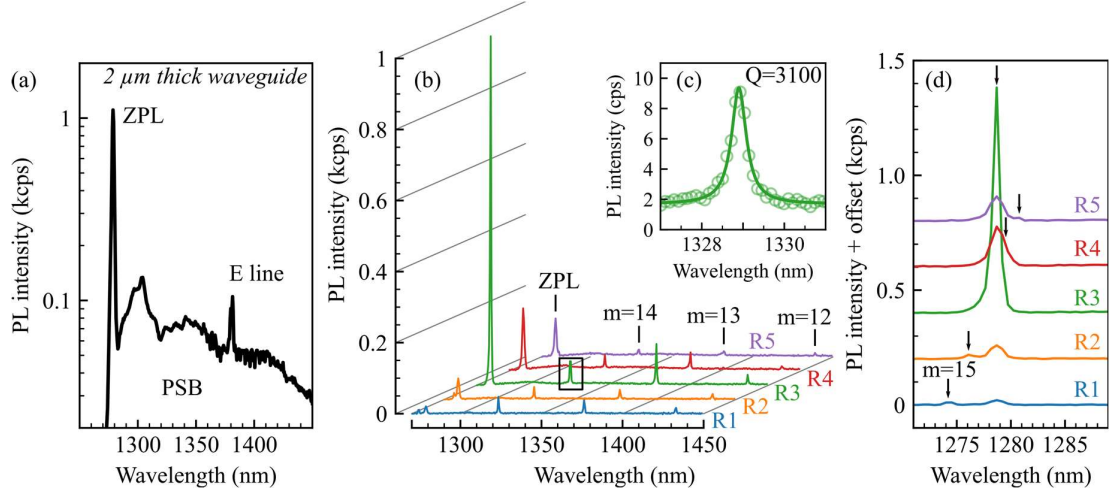

FIG. 2. PL spectra of G centers in (a) a 2 μm thick waveguide and (b) for all 5 microrings R1 to R5. (c) High-resolution spectrum of the $TE_{00,14}$ mode of R3. (d) Zoom on the ZPL and $TE_{00,15}$ mode.

The resonant modes slightly shift toward higher wavelengths as the diameter of the microring is increased. This allowed us to finely tune the $TE_{00,15}$ mode with the ZPL. The matching is obtained for R3, resulting in a remarkable enhancement of the ZPL intensity on the raw spectra (Fig. 2(b) and 2(d)). By normalizing the spectra with respect to the intensity of the PSB (see the discussion about the normalization procedure in the SI), we show that the ZPL amplitude is enhanced by a factor of $5.5 \pm 0.9$.

Time-resolved PL measurements were performed for the five microrings. The excitation laser had a repetition period of 60 ns and pulse duration of 300 ps. The ZPL was spectrally selected using a ±6 nm bandpass filter. A bi-exponential decay was obtained for all the microrings, as shown in Fig. 3. A short time $\tau_1 = 5.6$ ns and long time $\tau_2 = 26$ ns were measured on average, with respective standard deviations $\sigma_1 = 0.1$ ns and $\sigma_2 = 3$ ns on the series of measurements. The short time agrees well with previously reported values for ensembles of G centers[15]. Using a 1200 – 1250 nm band-pass filter to select the $TE_{00,16}$ mode of R3 at 1233 nm and reject G centers emission, a nearly mono-exponential decay with a characteristic time of 26 ns was obtained (see the supplementary Fig. S2). This result indicates that the long time $\tau_2$ is due to parasitic luminescent defects. In order to compare the counts rates of the G centers alone, we have assumed that the PL signal related to the parasitic emitters is the same for the five microrings under study. We normalized the histograms by the number of long-delay events between 40 and 55 ns, which are predominantly due to the parasitic emitters. For R3, the count rate of the G centers is enhanced by a factor ~4. This value is consistent with the cw PL measurements.

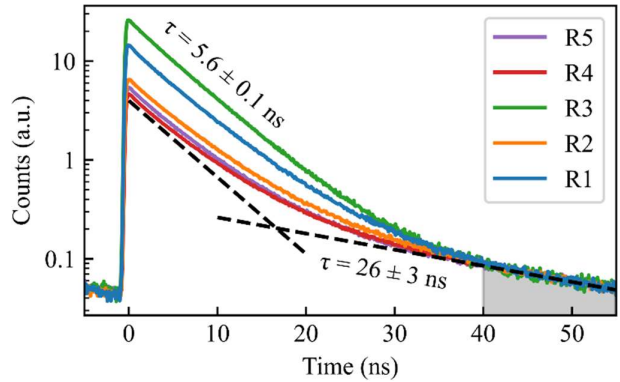

FIG. 3. Time-resolved PL measurements. The microrings were pumped using a pulsed laser with 10 µW average power at 532 nm, with a repetition period of 60 ns and pulse duration of 300 ps. The histograms are normalized by the total number of counts between 40 and 55 ns (grey area).

Despite this enhancement of zero-phonon emission, no sizeable decrease of the short decay was detected for R3. This result indicates that the dynamics of the excited state of the G centers is dominated by non-radiative processes.

In order to analyze these results more quantitatively, one has to consider how resonance modifies the photon emission process for a G-center. In the non-resonant case, the zero-phonon radiative transitions are emitted into a continuum of leaky modes at rate $\gamma\Gamma_0$, where $\Gamma_0$ denotes the zero-phonon SE rate of a G center in bulk silicon. It is well-known for tightly-confined structures such as nanowires that dielectric screening can inhibit SE into the leaky modes ($\gamma \ll 1$)[21,22]. This inhibition effect, which is at play in SOI microrings, has been explored using 3D FDTD simulations as a function of the position and dipole orientation of the G center (see the SI). An average inhibition factor $\bar{\gamma} \sim 0.4$ was calculated for the ensemble of G centers.

In the resonant case, the density of resonant modes adds to the density of leaky modes available for the radiative transitions. ZPL photons are therefore still emitted into the leaky modes at rate $\gamma\Gamma_0$, but also in the resonant mode at rate $F_p\Gamma_0$ where $F_p$ is the Purcell factor. The ZPL enhancement that is specifically observed for R3 is thus intrinsically due to the combination of Purcell enhancement in R3 and some inhibition of zero-phonon emission in the other microrings.

Now we estimate the average Purcell factor for the ensemble of G centers. In the case of a single emitter ideally coupled to a resonant mode[25], the Purcell factor reads $F_p = 3/4\pi^2 \, (\lambda/n)^3 \, Q/V$, where $\lambda$, $Q$ and $V$ are the wavelength, quality factor and volume of the mode, respectively. A volume $V \approx 7.2 \, (\lambda/n)^3$ is calculated from the intensity profile of the propagative mode, which gives $F_p \approx 33$. Considering the two-fold degeneracy of contra-propagative modes, the SE rate into resonant modes writes $2F_p\Gamma_0$. This factor must be averaged over a realistic spatial and spectral distribution of emitters. Using the SRIM software[26], the carbon implantation was simulated and fairly fitted to a Gaussian profile with a FWHM of 91 ± 1 nm. Assuming the distribution of the G centers $\rho(\mathbf{r})$ follows the carbon implantation profile, an averaging factor $\alpha_{spatial} = \iiint \rho(\mathbf{r})|\mathbf{E}(\mathbf{r})|^2 d\mathbf{r}/E_0^2 \iiint \rho(\mathbf{r})d\mathbf{r} \approx 0.62$ was calculated. As the ZPL and the mode have comparable FWHM, $F_p$ can be multiplied by a spectral averaging factor $\alpha_{spectral} = (1 + \Delta\lambda_{ZPL}/\Delta\lambda_{mode})^{-1} \approx 0.27$ with good approximation[27]. Finally, G centers dipoles can take three different orientations[28], so that a factor $\alpha_{orientation} = 1/3$ arises from the averaging of the overlap between the dipole and the local polarization of the resonant mode. In the end, we estimate that the zero-phonon SE rate of the G centers into the resonant mode is enhanced on average by a factor $\bar{F}_p = 2 \times \alpha_{spatial} \times \alpha_{spectral} \times \alpha_{orientation} \approx 3.7$ with respect to its value in the bulk.

To go one step further in the quantitative understanding of the increase of the PL intensity, one needs to consider how the photons are collected under resonant or non-resonant conditions. In the non-resonant case, ZPL photons emitted into the continuum of leaky modes of the microring are collected by the objective with efficiency $\eta_{leak}$. In the resonant case, ZPL photons are mostly emitted into the resonant mode. As Q is predominantly due to absorption losses by parasitic impurities, only a fraction $\sim Q_{abs}/Q_{rad}$ of this emission is scattered from the cavity, where $Q_{abs}$ is the absorption-limited Q, and $Q_{rad}$ is the radiation-limited Q including both bending losses and scattering by sidewall roughness. Additionally, this emission is collected by the microscope objective with efficiency $\eta_{mode}$. From these considerations, one can relate the intensity $I_{res}$ collected in the resonant case to the intensity $I_{off-res}$ collected in the non-resonant case:

$$I_{res} \approx \left(1 + \frac{\eta_{mode}}{\eta_{leak}} \frac{Q_{abs}}{Q_{rad}} \frac{\bar{F}_p}{\bar{\gamma}}\right) I_{off-res} \qquad (1)$$

For our configuration, $\eta_{mode} \sim 1.5\, \eta_{leak}$ (see the SI). Assuming as discussed above that propagation losses due to scattering are around 7dB/cm at 1.3 μm[20], bending losses and propagation losses have similar values, and $Q_{rad}$ would be around 30000. From Eqn. 1, $I_{res}/I_{off-res}$ would be close to 2.5. This estimate is a factor 2 smaller than the observed enhancement factor. It is likely that scattering-related losses are larger than state-of-the-art values for the present sample, which reduces $Q_{rad}$ and improves the collection efficiency of the part of the SE that has been injected into the cavity mode.

The increase of their zero-phonon SE rate should induce a decrease of the G centers lifetime by an average factor $\tau_{off-res}/\tau_{res} \approx 1 + \eta \xi \bar{F}_p$, where $\eta$ is the radiative yield of the emitters and $\xi \approx 15\%$ is their Debye-Waller factor. We define the threshold below which a lifetime decrease would be significant as $\tau_{th} = \tau_{off-r} - 3\sigma_1 = 5.3$ ns, where $\sigma_1 = 0.1$ ns is the standard deviation of the measured lifetimes defined previously. The same lifetime was measured when the ZPL was coupled or not to a resonant mode, meaning $\tau_{res} > \tau_{th}$ and therefore $\eta < \left(\frac{\tau_{off-res}}{\tau_{th}} - 1\right)/\xi \bar{F}_p \approx 0.1$. We conclude that the radiative yield of the present ensemble of G centers is less than 10%.

In conclusion, we have incorporated an ensemble of G centers in SOI microrings using ion implantation and standard nanofabrication, and observed a five-fold enhancement of the ZPL intensity. Our study highlights the detrimental role of parasitic defects that are produced besides G centers in heavily implanted silicon. Decreasing the implantation doses looks as a simple yet promising way to reduce residual absorption, resulting in higher Purcell factor and better extraction of the resonant mode. It might also improve the radiative yield of the emitters, by decreasing the density of non-radiative centers in their environment. Regarding G centers, recent studies at the single defect level have also shown that their radiative yield is intrinsically low[29]. From that respect, high-$\eta$ single defects such as W centers[10] and some carbon-related color centers[7] have a much higher potential for cavity quantum electrodynamics in SOI cavities. Using deterministic positioning techniques with nanometer scale resolution[29], a single defect with $\eta > 50\%$ (as in ref. 7) could be coupled nearly ideally with a resonant mode of a microring. For $Q \sim 3000$, the lifetime would be reduced by a factor $F = 6$ and the probability $P \approx 2\eta\xi F_p/(2\eta\xi F_p + 1)$ to emit a photon into the resonant mode after an excitation pulse would reach 70%. For (realistic) improved values of $Q$ around 20000, these values would be boosted up to $F = 34$ and $P = 97\%$. Such figures of merits would make single defects in SOI cavities appealing candidates for the development of deterministic single photon sources integrated on silicon chips.

See the Supplementary Information for more details about the finite-element simulations, measurement of the parasitic emitters lifetime, normalization procedure of the PL spectra and 3D FDTD simulations.

This work was supported by the French National Research Agency (ANR) through the OCTOPUS project, the Laboratoire d'Alliances Nanosciences-Energies du Futur (LANEF) labex and the French national 'Plan Quantique' through the QuantAlps Federation project. The authors acknowledge the assistance of the staff belonging to CEA Advanced Technological Platform (PTA).

AUTHOR DECLARATIONS

Conflict of interest

The authors have no conflict to disclose.

## DATA AVAILABILITY

The data that support the findings of this study are available from the corresponding author upon reasonable request.

## REFERENCES

[1] J.W. Silverstone, D. Bonneau, J.L. O'Brien, and M.G. Thompson, IEEE J. Sel. Top. Quantum Electron. **22**, 390 (2016).

[2] T. Rudolph, APL Photonics **2**, 030901 (2017).

[3] J. Wang, F. Sciarrino, A. Laing, and M.G. Thompson, Nat. Photonics **14**, 273 (2020).

[4] I. Aharonovich, D. Englund, and M. Toth, Nat. Photonics **10**, 631 (2016).

[5] Y. Baron, A. Durand, T. Herzig, M. Khoury, S. Pezzagna, J. Meijer, I. Robert-Philip, M. Abbarchi, J.-M. Hartmann, S. Reboh, J.-M. Gérard, V. Jacques, G. Cassabois, and A. Dréau, Appl. Phys. Lett. **121**, 084003 (2022).

[6] D.B. Higginbottom, A.T.K. Kurkjian, C. Chartrand, M. Kazemi, N.A. Brunelle, E.R. MacQuarrie, J.R. Klein, N.R. Lee-Hone, J. Stacho, M. Ruether, C. Bowness, L. Bergeron, A. DeAbreu, S.R. Harrigan, J. Kanaganayagam, D.W. Marsden, T.S. Richards, L.A. Stott, S. Roorda, K.J. Morse, M.L.W. Thewalt, and S. Simmons, Nature **607**, 266 (2022).

[7] W. Redjem, A. Durand, T. Herzig, A. Benali, S. Pezzagna, J.H. Meijer, A.Y. Kuznetsov, H.S. Nguyen, S. Cueff, J.-M. Gérard, I. Robert-Philip, B. Gil, D. Caliste, P. Pochet, M. Abbarchi, V. Jacques, A. Dréau, and G. Cassabois, Nat. Electron. **3**, 738 (2020).

[8] M. Hollenbach, Y. Berencén, U. Kentsch, M. Helm, and G.V. Astakhov, Opt. Express **28**, 26111 (2020).

[9] A. Durand, Y. Baron, W. Redjem, T. Herzig, A. Benali, S. Pezzagna, J.H. Meijer, A.Y. Kuznetsov, J.-M. Gérard, I. Robert-Philip, M. Abbarchi, V. Jacques, G. Cassabois, and A. Dréau, Phys. Rev. Lett. **126**, 083602 (2021).

[10] Y. Baron, A. Durand, P. Udvarhelyi, T. Herzig, M. Khoury, S. Pezzagna, J. Meijer, I. Robert-Philip, M. Abbarchi, J.-M. Hartmann, V. Mazzocchi, J.-M. Gérard, A. Gali, V. Jacques, G. Cassabois, and A. Dréau, ACS Photonics **9**, 2337 (2022).

[11] K. Murata, Y. Yasutake, K. Nittoh, S. Fukatsu, and K. Miki, AIP Adv. **1**, 032125 (2011).

[12] D.D. Berhanuddin, M.A. Lourenço, R.M. Gwilliam, and K.P. Homewood, Adv. Funct. Mater. **22**, 2709 (2012).

[13] C. Beaufils, W. Redjem, E. Rousseau, V. Jacques, A.Y. Kuznetsov, C. Raynaud, C. Voisin, A. Benali, T. Herzig, S. Pezzagna, J.H. Meijer, M. Abbarchi, and G. Cassabois, Phys. Rev. B **97**, 035303 (2018).

[14] L. Zhu, S. Yuan, C. Zeng, and J. Xia, Adv. Opt. Mater. **8**, 1901830 (2020).

[15] S.G. Cloutier, P.A. Kossyrev, and J. Xu, Nat. Mater. **4**, 887 (2005).

[16] M. Khoury, H. Quard, T. Herzig, J. Meijer, S. Pezzagna, S. Cueff, M. Abbarchi, H.S. Nguyen, N. Chauvin, and T. Wood, Adv. Opt. Mater. 2022, 2201295.

[17] A. Henry, B. Monemar, J.L. Lindström, T.D. Bestwick, and G.S. Oehrlein, J. Appl. Phys. **70**, 5597 (1991).

[18] G. Davies, Phys. Rep. **176**, 83 (1989).

[19] W. Bogaerts, P. De Heyn, T. Van Vaerenbergh, K. De Vos, S. Kumar Selvaraja, T. Claes, P. Dumon, P. Bienstman, D. Van Thourhout, and R. Baets, Laser Photonics Rev. **6**, 47 (2012).

[20] S. Feng, K. Shang, J.T. Bovington, R. Wu, B. Guan, K.-T. Cheng, J.E. Bowers, and S.J.B. Yoo, Opt. Express **23**, 25653 (2015).

[21] Q. Wilmart, S. Brision, J.-M. Hartmann, A. Myko, K. Ribaud, C. Petit-Etienne, L. Youssef, D. Fowler, B. Charbonnier, C. Sciancalepore, E. Pargon, S. Bernabé, and B. Szelag, J. Light. Technol. **39**, 532 (2021).

[22] J.K. Doylend, P.E. Jessop, and A.P. Knights, Opt. Express **19**, 14913 (2011).

[23] J.P. Zhang, D.Y. Chu, S.L. Wu, S.T. Ho, W.G. Bi, C.W. Tu, and R.C. Tiberio, Phys. Rev. Lett. **75**, 2678 (1995).

[24] J. Bleuse, J. Claudon, M. Creasey, Nitin.S. Malik, J.-M. Gérard, I. Maksymov, J.-P. Hugonin, and P. Lalanne, Phys. Rev. Lett. **106**, 103601 (2011).

[25] Monochromatic emitter perfectly matching the mode, located at an antinode, with its dipole parallel to the electric field.

[26] See http://www.srim.org/ for information.

[27] J.-M. Gérard, in *Single Quantum Dots: Fundamentals, Applications and New Concepts,* ed. Michler P, Springer, Berlin Heidelberg, Topics in Applied Physics 90, 269 (2003)[28] P. Udvarhelyi, B. Somogyi, G. Thiering, and A. Gali, Phys. Rev. Lett. **127**, 196402 (2021).

[29] Durand et al., in preparation.

[30] M. Hollenbach, N. Klingner, N.S. Jagtap, L. Bischoff, C. Fowley, U. Kentsch, G. Hlawacek, A. Erbe, N.V. Abrosimov, M. Helm, Y. Berencén, and G.V. Astakhov, arXiv:2204.13173 (2022)

# Cavity-enhanced zero-phonon emission from an ensemble of G centers in a silicon-on-insulator microring

## Supplementary Information

B. Lefaucher[1], J.-B. Jager[1], V. Calvo[1], A. Durand[2], Y. Baron[2], F. Cache[2], V. Jacques[2], I. Robert-Philip[2], G. Cassabois[2], T. Herzig[3], J. Meijer[3], S. Pezzagna[3], M. Khoury[4], M. Abbarchi[4], A. Dréau[2], J.-M. Gérard[1,a)].

[1] *Université Grenoble Alpes, CEA, Grenoble INP, IRIG, PHELIQS, Grenoble, France.*

[2] *Laboratoire Charles Coulomb, Université de Montpellier and CNRS, Montpellier, France.*

[3] *Division of Applied Quantum Systems, Felix-Bloch Institute for Solid-State Physics, University Leipzig, Leipzig, Germany.*

[4] *CNRS, Aix-Marseille Université, Centrale Marseille, IM2NP, UMR 7334, Marseille, France.*

[a)] Author to whom correspondence should be addressed : jean-michel.gerard@cea.fr

CONTENTS

FIELD MAP OF THE TE$_{00}$ GUIDED MODE

Fig. S1 shows a vectorial cross-section field map of the TE$_{00}$ guided mode at 1.28 μm of a curved waveguide with 1.21 μm radius, and same structure as R3. The field map has been derived from an exact resolution of Maxwell's equations in cylindrical coordinates, using the finite-element solver FEMSIM from SYNOPSYS. The volume of the mode is defined as:

$$V = \frac{\iiint n^2(\mathbf{r})\,|\mathbf{E}(\mathbf{r})|^2 d\mathbf{r}}{n^2|\mathbf{E}_{max}|^2} \quad (1)$$

where $n(\mathbf{r})$ and $|\mathbf{E}(\mathbf{r})|^2$ are the local refractive index and intensity of the electromagnetic field, and $|\mathbf{E}_{max}|^2$ is the maximum field intensity. A volume $V = 7.2\ (\lambda/n)^3$ was calculated.

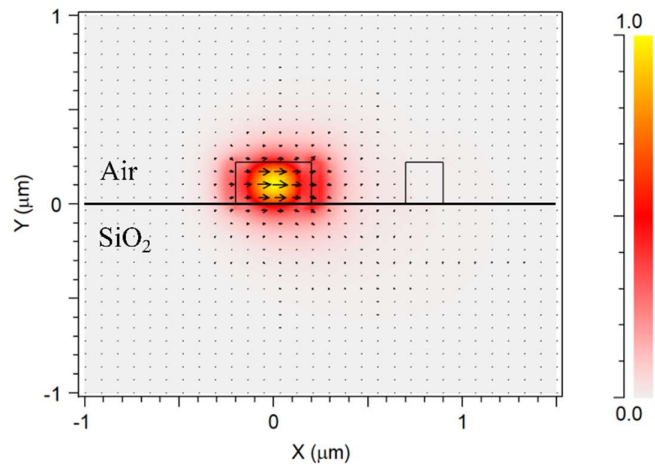

FIG. S1. The vectorial electric field map and the field intensity color map are shown over the cross-section of the curved waveguide for the TE$_{00}$ fundamental guided mode. The linear color scale is normalized to one at the antinode of the mode.

# GROUP INDEX, FREE SPECTRAL RANGE AND AZIMUTHAL INDEX OF RESONANT MODES FOR MICRORING R3

The effective index $n_{eff}$ of the TE$_{00}$ guided mode was calculated for several wavelengths between 1.25 and 1.45 µm, using the index values of silicon $n_{Si}$ at 10 K[1]. The group index of the mode, defined as $n_g = n_{eff} - \lambda\, \partial n_{eff}/\partial \lambda$, was then calculated for several wavelengths. The results are summarized in Table 1. A group index of 4.527 is obtained at 1.35 µm, from which the Free Spectral Range (FSR) can be calculated according to :

$$FSR = \frac{\lambda^2}{\pi D n_g} \qquad (2)$$

where $D$ is the effective diameter of the mode, roughly equal to the average diameter of the microring as we can see from the intensity profile of the mode in Fig. S1. A FSR of 52.8 nm is calculated, in excellent agreement with the experimental results.

Table 1. Effective and group indices of the TE$_{00}$ mode at 10 K.

| Wavelength (µm) | $n_{Si}$ at 10 K | $n_{eff}$ | $n_g$ |
|---|---|---|---|
| 1.250 | 3.483 | 2.565 |  |
| 1.275 | 3.479 | 2.526 | 4.488 |
| 1.300 | 3.475 | 2.488 | 4.500 |
| 1.325 | 3.472 | 2.449 | 4.513 |
| 1.350 | 3.469 | 2.410 | 4.527 |
| 1.375 | 3.466 | 2.376 | 4.541 |
| 1.400 | 3.463 | 2.331 | 4.555 |
| 1.425 | 3.460 | 2.291 | 4.568 |
| 1.450 | 3.458 | 2.251 |  |

Knowing the effective index, we can also deduce the azimuthal quantum number m of the resonant modes that are observed on the PL spectrum of the microrings, which is given in good approximation by :

$$m \approx 2\pi \bar{R}\, \frac{n_{eff}(\lambda_m)}{\lambda_m} \qquad (3)$$

where $\bar{R}$ is the average radius of the ring. Table 2 shows that the integer value of m can be estimated without ambiguity for all modes.

Table 2. Estimated azimuthal index m for the resonant modes of microring R3

| Measured $\lambda_m$ (µm) | 1.2791 | 1.3278 | 1.3808 | 1.4372 |
|---|---|---|---|---|
| Estimated m | 15.01 | 14.02 | 13.02 | 12.04 |

LIFETIME MEASUREMENT OF THE PARASITIC EMITTERS

The presence of parasitic emitters is revealed by the excitation of the $TE_{00,16}$ mode for all the microrings on photoluminescence (PL) spectra. Fig. S2a,b shows an example for R3. In order to measure the lifetime of the emitters coupled to the $TE_{00,16}$ mode, a 1200-1250 nm band-pass was used to filter out the G centers emission (Fig. S2c). The resulting decay curve is shown in Fig. S2d. A bi-exponential decay is obtained with 90% of the collected photons associated with a characteristic time $\tau_2 = 26$ ns, and 10% with a time $\tau_1 = 5.5$ ns attributed to residual G centers emission passing through the filter.

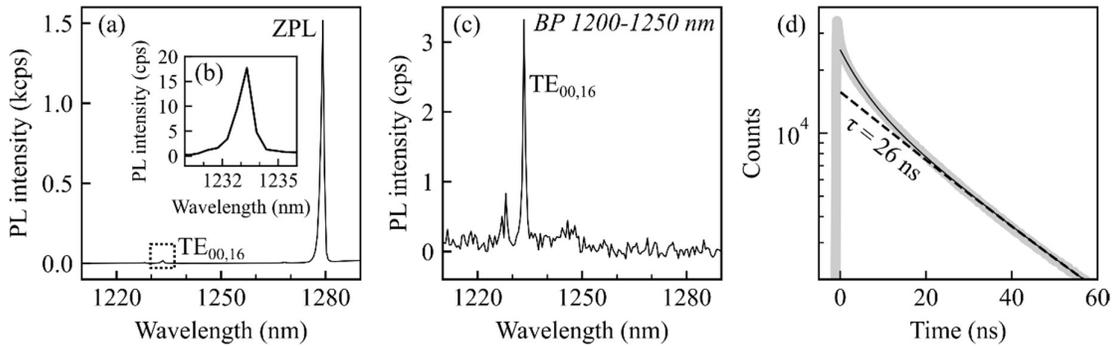

FIG. S2. Lifetime measurement of the parasitic emitters. (a) PL spectrum of R3. (b) Zoom on the $TE_{00,16}$ mode. (c) PL spectrum obtained in the same excitation conditions with a band pass filter. (d) Decay curve obtained using the band-pass filter.

## CALCULATION OF THE ENHANCEMENT FACTOR FROM PL SPECTRA

In order to compare quantitatively the intensity of the ZPL for different microrings, one must compensate for possible fluctuations of electron-hole pairs injection between different measurements. Since experiments are performed in the linear regime, and since the G centers lifetime is observed to remain constant, identical pumping conditions should lead to the same PL intensity for off-resonant phonon-assisted emission. We show in Fig.S3a the PL spectra of the five microrings, after normalization by the intensity integrated between 1290 and 1320 nm. Using this procedure, the intensity of the ZPL thus becomes similar for all the off-resonant cavities, as shown in Fig. S3b.

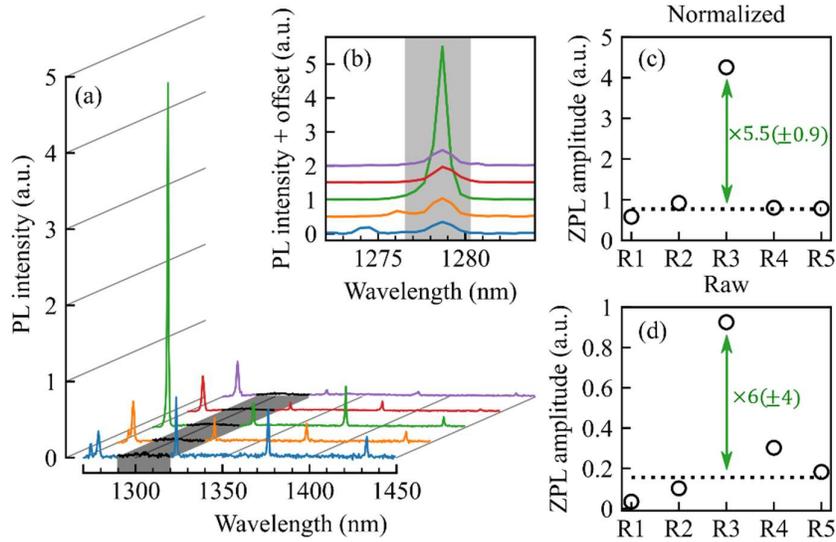

FIG. S3. Definition of an enhancement factor. (a) PL spectra normalized with respect to the intensity integrated between 1290 and 1320 nm (grey area). (b) Comparison of the ZPL intensities after normalization. (c) Comparison of the integrated ZPL intensities of the five microrings, after normalization. The ZPL intensity has been integrated over the spectral range marked by the grey area in (b). The dotted line indicates the average value for the four off-resonant rings. (d) Same as (c), for raw, unnormalized PL spectra.

Fig. S3c shows the amplitude of the ZPL for each microring, corresponding to the intensity integrated in the grey area of Fig. S3b. The dotted line indicates the average amplitude of the off-resonant rings. An enhancement factor of 5.5±0.9 is thus defined by comparison with the resonant cavity R3. Fig. S3d shows the amplitude values obtained from the raw spectra. Although the average PL intensity enhancement factor at resonance is also close to 5, the fluctuations of the ZPL intensity between off resonant rings leads to a large uncertainty on its estimate ($\frac{I_{res}}{I_{off-res}} \approx 6 \pm 4$). This highlights the interest of the normalization procedure.

## 3D FDTD SIMULATIONS

We have used the FULLWAVE software from Synopsys Corp to estimate several important parameters through 3D finite-differences in the time domain (FDTD) simulations.

**1. The intrinsic quality factor Q of whispering gallery modes**, only limited by bending losses, has been calculated for the resonant modes of microring R3. In a first numerical experiment, an emitter generates a broadband flash of light at t=0. The frequency and linewidth (hence Q) of the resonant modes are obtained from a Fourier transform of the field that is still trapped in the cavity after a sufficiently long time. Once resonant mode frequencies are known, one can also simulate a cavity ring-down experiment. A single mode is launched in the cavity at t=0; by monitoring cavity losses and the exponential decay of the stored electromagnetic energy, one gets access to the cavity storage time and Q. This method has been used to cross-check estimates obtained by the first method, especially for the mode of order m=15. Values that are obtained for Q by both methods agree within typically 10% as shown in Table 3.

Table 3. Estimation of the quality factor of the $TE_{00m}$ modes.

| Mode order m | 11 | 12 | 13 | 14 | 15 |
|---|---|---|---|---|---|
| Q (spectral analysis) | 2400 | 7400 | 17200 | 39000 | 76000 |
| Q (ring-down) |  | 7900 |  |  | 82000 |

**2. The collection efficiency $\eta_{mode}$** corresponds to the proportion of radiation escaping from a resonant mode that is captured by the microscope objective of our experimental set-up. In the numerical simulations, we launch a single resonant mode inside the microring. Once a stationary regime is reached at the edges of the simulation box, we stop the simulation and calculate the radiation pattern in the far field using a standard procedure. $\eta_{mode}$ is given by the ratio of the power collected in a NA=0.85 centered at normal incidence, by the total radiated power. Importantly, we assume that the polar distribution of the far field is the same for an ideal ring and for a real ring displaying sidewall roughness. This is reasonable since the sidewall corrugations that are introduced by reactive ion-etching are mostly made of vertical striations, that are not likely to modify the polar distribution of the far field through scattering[2]. A collection efficiency $\eta_{mode} \approx 15\%$ was calculated.

**3. Spontaneous emission (SE) rate $\gamma \Gamma_0$ for out-of-resonance emitters**. When the ZPL is not in resonance with one of the TE$_{00m}$ resonant modes, zero-phonon emission is only coupled to leaky modes of the microring. The $\gamma$ factor denotes the SE rate into the leaky modes normalized to the SE rate $\Gamma_0$ in bulk silicon. In the numerical simulation, we place in the ring a linear monochromatic dipole emitting at a wavelength that is few nm larger than the one of mode TE$_{00,15}$. The dipole emits in cw mode, after a slow ramp-up (necessary to avoid any excitation of resonant modes). The SE rate enhancement/inhibition factor $\gamma$ is given by the SE rate in the ring resonator divided by the emission rate of the same dipole in bulk silicon.

We calculate $\gamma$ for 9 different positions in the cross-section of the waveguide and various dipole orientations, as shown in Fig. S4. For G centers in bulk silicon, the dipole can take one of the three orientations XY=[110], XZ=[101] and YZ=[011], depending on the location of the Si auto-interstitial bridging the two carbon substitutional atoms[3]. In ring resonators, $\gamma$ depends upon the relative orientation of the dipole with respect to the radial direction, defined by angle $\phi$. For simulations, we can thus choose X as radial direction at the location of the dipole, and consider all possible relative orientations. More precisely:

- For the horizontal dipole, we calculate $\gamma$ and $\eta_{leak}$ for radial (X=[100]) and orthoradial (Y=[010]) orientations, and get results for other orientations using :

$$\gamma(\phi) = \gamma_X \cos^2 \phi + \gamma_Y \sin^2 \phi \qquad (4)$$

- Similarly, for the two other dipoles, we consider first XZ and YZ orientations, and get results for other orientations by using a formula similar to (4).

Let us mention that Eqn. (4) is justified because leaky modes can be separated in two families of modes, with radial or orthoradial polarizations with respect to the ring. Its validity has also been confirmed by running the simulation for a dipole such as XY. Excellent agreement with estimates obtained using (4) is observed for all nine positions considered for the dipole.

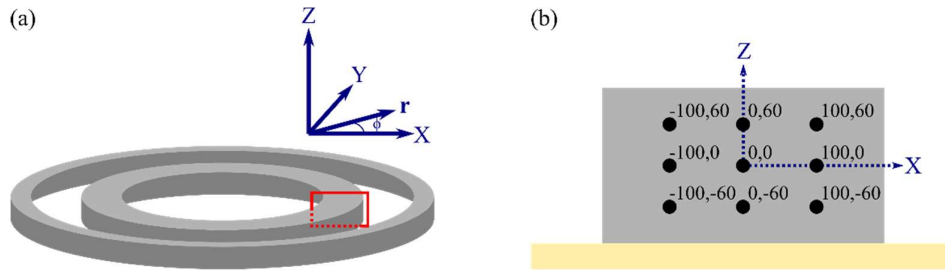

Fig. S4. Configuration of the FDTD simulation. (a) Representation of the cavity. (b) Cross-section of the microring indicating the 9 locations of the dipole in nm for the simulations.

Assuming now that the three orientations of the optical dipole are equally probable for G centers (which is true in bulk material for symmetry reasons, and a good approximation in SOI films according to recent experimental results[4] the averaging over all possible dipole orientations leads to the following average values for $\gamma$:

$$\langle \gamma \rangle_\phi = \frac{1}{6}(\gamma_X + \gamma_Y + 2\gamma_{XZ} + 2\gamma_{YZ}) \tag{5}$$

Table 4 shows the values obtained for $\gamma$, for each position of the dipole, from which the $\langle \gamma \rangle_\phi$ values were calculated. Note that the values for the XY dipole are given here as a validity check for (4), but are not necessary for estimating $\langle \gamma \rangle_\phi$ from equation (5).

Table 4. Normalized SE rate $\gamma$ for X, Y, XY, YZ, XZ dipoles for various locations in the cross-section of the ring.

| Location (nm) | $\gamma_X$ | $\gamma_Y$ | $\gamma_{XY}$ | $\gamma_{XZ}$ | $\gamma_{YZ}$ | $\langle \gamma \rangle_\phi$ |
|---|---|---|---|---|---|---|
| (0,0) | 0.09 | 0.70 | 0.40 | 0.13 | 0.45 | 0.33 |
| (0,60) | 0.79 | 0.69 | 0.74 | 0.35 | 0.40 | 0.50 |
| (0,-60) | 0.67 | 1.09 | 0.88 | 0.51 | 0.60 | 0.66 |
| (100,0) | 0.13 | 0.35 | 0.24 | 0.31 | 0.46 | 0.33 |
| (100,60) | 0.46 | 0.65 | 0.55 | 0.15 | 0.44 | 0.38 |
| (100,-60) | 0.39 | 0.63 | 0.51 | 0.58 | 0.47 | 0.52 |
| (-100,0) | 0.09 | 0.35 | 0.22 | 0.21 | 0.31 | 0.25 |
| (-100,60) | 0.49 | 0.49 | 0.49 | 0.45 | 0.30 | 0.41 |
| (-100,-60) | 0.42 | 0.59 | 0.51 | 0.11 | 0.36 | 0.32 |

In order to get an average value $\bar{\gamma}$ over a realistic distribution of dipoles, the $\langle \gamma \rangle_\phi$ values at various locations must be weighted by the local density of the dipoles in the microring. The PL maps indicate that the density of emitters is uniform in the XY plane, while SRIM simulations show a Gaussian distribution of the carbon atoms along the Z axis with a FWHM of 91 nm. We can thus roughly estimate that 60% of the emitters are located around the center of layer, 20% in the top third and 20% in the bottom third, and calculate $\bar{\gamma}$ according to:

$$\bar{\gamma} = 60\% \times \frac{1}{3} \sum_{\substack{z=0 \\ x=0,\pm 100 \text{ nm}}} \langle \gamma_{x,z} \rangle_\phi + 20\% \times \frac{1}{6} \sum_{\substack{z=\pm 60 \text{ nm} \\ x=0,\pm 100 \text{ nm}}} \langle \gamma_{x,z} \rangle_\phi \tag{6}$$

We obtain $\bar{\gamma} \approx 0.37$.

**4. Collection efficiency $\eta_{leak}$ for out-of-resonance emitters.** The power emitted in a solid angle corresponding to NA=0.85 around normal incidence was monitored for each dipole in the simulation. Depending on the relative orientation of the dipole with respect to the radial direction, a fraction $\eta_{leak}$ of the normalized SE rate $\gamma$ is emitted in the solid angle. A value of $\eta_{leak}\gamma$ is thus obtained for each dipole from the simulation, corresponding to the intensity collected by the microscope objective with respect to the total emitted intensity for our experimental configuration. The results are summarized in Table 5.

We can thus calculate the total intensity $\overline{\gamma\eta}_{leak}$ collected from a realistic distribution of out-of-resonance dipoles using the same averaging procedure as for $\bar{\gamma}$. We obtain $\overline{\gamma\eta}_{leak} \approx 0.040$. An average collection efficiency can thus be defined for out-of-resonance emitters as $\bar{\eta}_{leak} = \overline{\gamma\eta}_{leak}/\bar{\gamma} \approx 11\%$.

Table 5. Normalized intensity $\eta_{leak}\gamma$ collected in NA=0.85 for X, Y, XY, YZ, XZ dipoles for various locations in the cross-section of the ring.

| Location (nm) | $\gamma_X\eta_{leak,X}$ | $\gamma_Y\eta_{leak,Y}$ | $\gamma_{XY}\eta_{leak,XY}$ | $\gamma_{XZ}\eta_{leak,XZ}$ | $\gamma_{YZ}\eta_{leak,YZ}$ | $\langle\gamma\eta_{leak}\rangle_\phi$ |
|---|---|---|---|---|---|---|
| (0,0) | 0.019 | 0.095 | 0.057 | 0.013 | 0.054 | 0.041 |
| (0,60) | 0.051 | 0.080 | 0.066 | 0.027 | 0.043 | 0.045 |
| (0,-60) | 0.024 | 0.260 | 0.140 | 0.017 | 0.135 | 0.098 |
| (100,0) | 0.016 | 0.036 | 0.026 | 0.010 | 0.029 | 0.022 |
| (100,60) | 0.035 | 0.075 | 0.054 | 0.015 | 0.040 | 0.037 |
| (100,-60) | 0.018 | 0.144 | 0.052 | 0.023 | 0.080 | 0.061 |
| (-100,0) | 0.015 | 0.045 | 0.030 | 0.024 | 0.031 | 0.028 |
| (-100,60) | 0.031 | 0.090 | 0.060 | 0.025 | 0.050 | 0.045 |
| (-100,-60) | 0.015 | 0.115 | 0.065 | 0.009 | 0.065 | 0.046 |

REFERENCES

[1] B.J. Frey, D.B. Leviton, and T.J. Madison, Proc SPIE **6273**, (2006).

[2] W. Bogaerts, P. De Heyn, T. Van Vaerenbergh, K. De Vos, S. Kumar Selvaraja, T. Claes, P. Dumon, P. Bienstman, D. Van Thourhout, and R. Baets, Laser Photonics Rev. **6**, 47 (2012).

[3] P. Udvarhelyi, B. Somogyi, G. Thiering, and A. Gali, Phys. Rev. Lett. **127**, 196402 (2021).

[4] Durand et al., in preparation.